\documentclass[runningheads]{llncs}
\usepackage[a-1b]{pdfx}
\usepackage{graphicx}
\usepackage[all, cmtip]{xy}
\usepackage{mathtools}
\usepackage{amsfonts}
\usepackage{amssymb}
\usepackage{dsfont}
\usepackage{amsmath}
\usepackage{tikz}
\usepackage{tikz-cd}
\usepackage{hyperref}
\usepackage{subfigure}
\usepackage{enumitem}

\begin{document}
%
%
\title{A Formal Category Theoretical Framework for Multi-Model Data Transformations}
\titlerunning{A Category Theoretical Framework for Multi-Model Data Transformations}
%
\author{Valter Uotila \and Jiaheng Lu}
%
%
\institute{Unified Database Management Systems, University of Helsinki, Helsinki, Finland \\
\email{\{valter.uotila,jiaheng.lu\}@helsinki.fi}
}
\maketitle              
\begin{abstract}
Data integration and migration processes in polystores and multi-model database management systems highly benefit from data and schema transformations. Rigorous modeling of transformations is a complex problem. The data and schema transformation field is scattered with multiple different transformation frameworks, tools, and mappings. These are usually domain-specific and lack solid theoretical foundations. Our first goal is to define category theoretical foundations for relational, graph, and hierarchical data models and instances. Each data instance is represented as a category theoretical mapping called a functor. We formalize data and schema transformations as Kan lifts utilizing the functorial representation for the instances. A Kan lift is a category theoretical construction consisting of two mappings satisfying the certain universal property. In this work, the two mappings correspond to schema transformation and data transformation.

\end{abstract}

\section{Introduction}

The biggest success stories in database theory are the relational model and relational algebra. Codd's theory \cite{10.1145/362384.362685} on relational databases has had an incomprehensible huge impact on database theory and applications. More formal and theoretical treatment of polystores and multi-model databases would make it possible us to repeat this success story in polystores and multi-model databases. A solid mathematical foundation would highly benefit their research and industry. Besides, to standardize the existing techniques and systems, a rigorous formulation is crucial.

Polystores and multi-model databases are a solution to the problem of handling a variety of data \cite{7761636,10.1007/978-3-030-33752-0_10,DBLP:conf/cikm/LuHC18}. Native graph, document, key-value, and column databases have reached the point where they are competitive alternatives for relational databases especially in the cases when we perform a lot of read- and write-operations and heavy data analysis tasks. Since ML and AI are relying on massive amounts of data, NoSQL databases have gained attention.

Undoubtedly, polystores and multi-model databases are more complicated systems than ordinary relational databases since they subsume relational databases. The theory and language describing the systems have to evolve along with the systems which are gradually becoming more complex. But this should not mean that the theory and languages become more complex for end-users or even for database administrators and architects. Different databases have their own theoretical foundations and query languages that are not automatically compatible at a practical or theoretical level. This creates a huge challenge that we are tackling from the theoretical perspective.

Data and schema transformations form a significant part of the data integration and migration problems \cite{DBLP:conf/hpec/DziedzicES16}. For example, the transformations might be needed at any point during the development of ML and AI solutions where databases are a part of the process. Initially, importing data requires transformations. Data integration between the databases can require multiple transformation-based views between the participating databases. Sometimes the most efficient solution is to materialize the transformed data. When the amount of data grows, the transformation systems need to be able to adapt for the growth. Thus monotonicity and temporality aspects of transformations are important to take into account. Eventually, the data require transformations before it can fit ML and AI models. For example, ML and AI models can use a knowledge graph approach but the data is stored in a relational database. The same transformation problems are also apparent for polystores and multi-model databases. 

Often these transformations lack formal treatment. Daimler et al. \cite{Daimler_Wisnesky_2020} argue that informal data transformations are harmful. This is one of the challenges we are addressing in this work. The language, which is proved to be capable of capturing highly complex structures with a compact notation, is category theory. Liu et al. \cite{Liu2018MultimodelDM} visioned that the foundations of multi-model databases could be built on category theory because relational algebra's expressiveness is not powerful enough. We argue that the same applies to polystores. Our contributions include
\begin{itemize}
    \item continuing previous research connecting category theory and database theory,
    \item formalizing graph and hierarchical models and instances in terms of category theory, and
    \item formalizing data transformations in polystores and multi-model databases as a solution to a category theoretical lifting problem.
\end{itemize}

Informally category can be thought of as a graph or a network with a certain additional structure. The additional structure is usually easy to find from computer science and database applications. If our goal is to express database theory precisely, it does not make sense to use only graphs because we can do modeling much better with categories.

In this work, we are often mentioning ''schema''. By schema, we do not only mean the conventional relational schema but a larger piece of information that contains any constraint related to a model. Also, the information about the model is part of the schema. Although modern NoSQL data is often referred to as schemaless, the data always have some constraints which we include in a schema in this context.

\subsection{Related work}

There are influential transformation frameworks but only a few of them are developed formally. SQLGraph \cite{Sun_Fokoue_Srinivas_Kementsietsidis_Hu_Xie_2015} is a system, which translates graph databases to relational databases. It utilizes hashing and the fact that the modern relational databases natively support JSON. A framework of converting relational databases to graph databases by Virgilio et al. \cite{De_Virgilio_Maccioni_Torlone_2013} utilizes schema paths. Das et al. \cite{Das_Srinivasan_Perry_Chong_Banerjee_2014} have developed a framework that creates RDF-view for property graph data in Oracle databases. All of these transformations are considered from a domain-specific and practical perspective although we identify that they have characteristical features which could be theoretically modeled and unified. 

Jananthan et al. \cite{8258298} propose associative algebra as a mathematical foundation for polystores. Leclercq et al. \cite{10.1145/3216122.3216152} built foundations of polystores on the tensor-based data model. Liu et al. \cite{Liu2018MultimodelDM} visioned that the foundations of multi-model databases could be built on category theory and we continue this work for polystores and multi-model databases.

There has been relatively much research on applying category theory to database theory. Our approach is highly influenced by David Spivak \cite{Spivak2013DatabaseQA,spivak2014category}. As he points out in \cite{Spivak2013DatabaseQA}, the category theoretical database research can be divided into two schools: category-based \cite{functorialDataMigration} and sketch-based \cite{Kadish1994AlgebraicG}. A sketch \cite{wells2012} is a category with certain limit objects. Our position is category-based. 

Besides work on database theory, category theory has been applied widely in computer science. Some of the most interesting and recent applications are programming languages (foundations of many functional programming languages), machine learning \cite{DBLP:journals/corr/abs-2103-01931,8785665}, automata learning \cite{DBLP:journals/corr/abs-2001-05786}, natural language processing (DisCoCat \cite{DBLP:journals/corr/abs-1003-4394}), and quantum computing and mechanics \cite{abramsky2008categorical,Coecke2011}. Applied category theory has its annually organized conference called ACT (Applied Category Theory).
\section{Prerequisites}
\subsection{Categories}
Category theory is a relatively new field of mathematics. Saunders MacLane and Samuel Eilenberg introduced categories, functors, and natural transformations in the mid-1940s as a ''meta-mathematical'' tool to study algebraic topology. MacLane \cite{lane1998categories} is the standard introduction to the topic. Other good introduction from mathematical perspective is \cite{riehl2017category} and from computer science perspective \cite{spivak2014category,wells2012}.

\begin{definition}[Category]\label{def:category}
A category $\mathsf{C}$ consists of a collection of objects denoted by $Obj(\mathsf{C})$ and a collection of morphisms denoted by $Hom(\mathsf{C})$.  For each morphism $f \in Hom(\mathsf{C})$ there exists an object $A \in Obj(\mathsf{C})$ that is a domain of $f$ and an object $B \in Obj(\mathsf{C})$ that is a target of $f$. In this case we denote $f \colon A \to B$. We require that all the defined compositions of morphisms are included in $\mathsf{C}$: if $f\colon A \to B \in Hom(\mathsf{C})$ and $g \colon B \to C \in Hom(\mathsf{C})$ are morphisms, then the composition $g \circ f \in Hom(\mathsf{C})$ is defined and $g \circ f \colon A \to C$ is a morphism. Also, we assume that the composition operation is associative and that for every object $A \in Obj(\mathsf{C})$ there exists an identity morphism $\text{id}_{A} \colon A \to A$ so that $f \circ \text{id}_{A} = f$ and $\text{id}_{A} \circ f = f$ whenever the composition is defined.
\end{definition}

\begin{figure}
    \centering
    \includegraphics[scale=0.7]{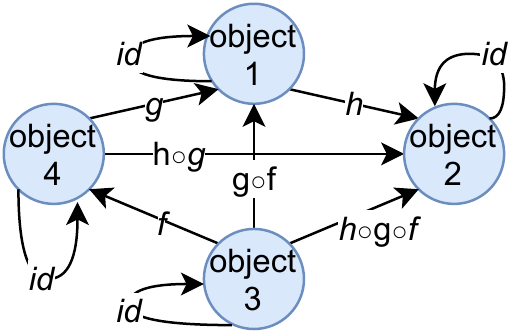}
    \caption{A simple four object category with three non-trivial morphisms $f$, $g$ and $h$ and identities. In this case all the compositions of morphisms are drawn.}
    \label{fig:category}
\end{figure}

See \autoref{fig:category} as a simple example of a category. In this work \textsf{sans serif} font always indicates a category. We follow the standard notation of category theory literature that is used, for example, in \cite{riehl2017category}.
One of the most important categories is the category $\mathsf{Set}$ whose objects are sets and morphisms are functions between the sets. The composition operation of the morphisms is the composition of functions.

\subsection{Functors}
In science and mathematics, we often have functions or mappings which respect the underlying structures. Next, we define a structure-preserving mapping for categories. The mapping is called a functor.
\begin{definition}[Functor]\label{def:functor}
Assume $\mathsf{C}, \mathsf{D}$ are categories. A functor $F \colon \mathsf{C} \to \mathsf{D}$ is defined so that
\begin{itemize}[noitemsep, topsep=1pt]
\item for every object $c$ in the category $\mathsf{C}$, $F(c)$ is an object in the category $\mathsf{D}$ and
\item for every morphism $f \colon c \to d$ in $\mathsf{C}$, it holds that $F(f) \colon F(c) \to F(d)$ is a morphism in $\mathsf{D}$.
\end{itemize}
Besides, we assume that following axioms hold:
\begin{itemize}[noitemsep, topsep=1pt]
    \item For every object $c \in \mathsf{C}$ it holds that $F(\text{id}_c) = \text{id}_{F(c)}$ and
    \item if the composition $f \circ g$ is defined, then $F(f \circ g) = F(f) \circ F(g)$.
\end{itemize}
If every morphism in the category $\mathsf{D}$ has a preimage in the category $\mathsf{C}$, we call the functor $F$ full.
\end{definition}
See \autoref{fig:category_functor_natural_transformation_examples} (a) as an example of functor between two simple categories. The fact that a functor preserves the structure of a category is apparent in the example.

\subsection{Natural transformations}
The idea behind structure-preserving mappings is so fundamental that we can study what it means to preserve a structure of structure-preserving mappings. The category theoretical notion for this is called a natural transformation. We follow a convention from category theory and denote a natural transformation by ''$\Rightarrow$''-arrow.

\begin{definition}[Natural transformation]\label{def:naturalTransformation}
Assume $F, G \colon \mathsf{C} \Rightarrow \mathsf{D}$ are functors. A natural transformation $\alpha \colon F \Rightarrow G$ contains the following information: For each $c \in \mathsf{C}$ is associated a \textit{component} of the natural transformation $\alpha_c \colon F(c) \to G(c)$. This component is a morphism in $\mathsf{D}$ so that the following diagram commutes for any morphism $f \colon c \to d$ in $\mathsf{C}$
\begin{displaymath}
\xymatrix{
F(c) \ar[d]_{F(f)} \ar[r]^{\alpha_c} & G(c) \ar[d]^{G(f)} \\
F(d) \ar[r]_{\alpha_d} & G(d)
}
\end{displaymath}
In equational format commuting means that $G(f) \circ \alpha_c = \alpha_d \circ F(f)$.
\end{definition}

\begin{figure}
\centering
\subfigure[]{\includegraphics[width = 0.49\textwidth]{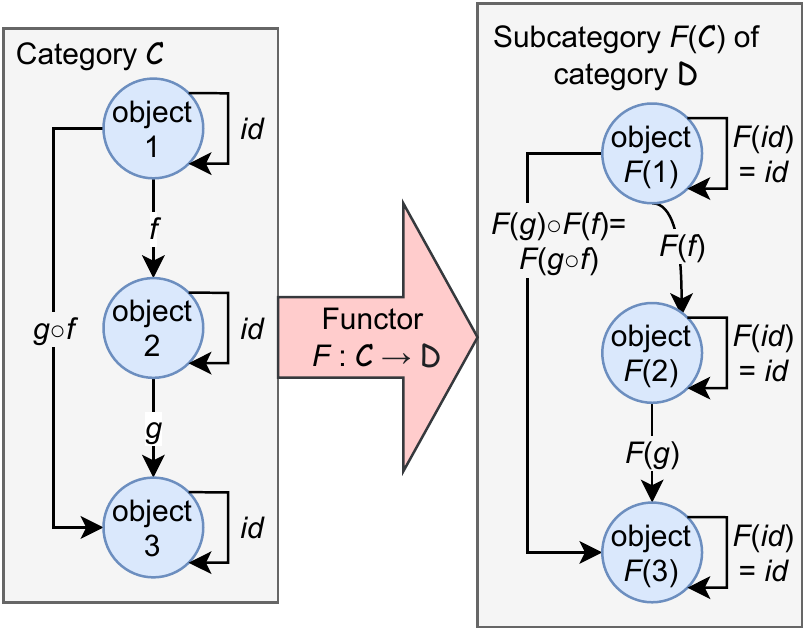}}
\subfigure[]{\includegraphics[scale = 0.59]{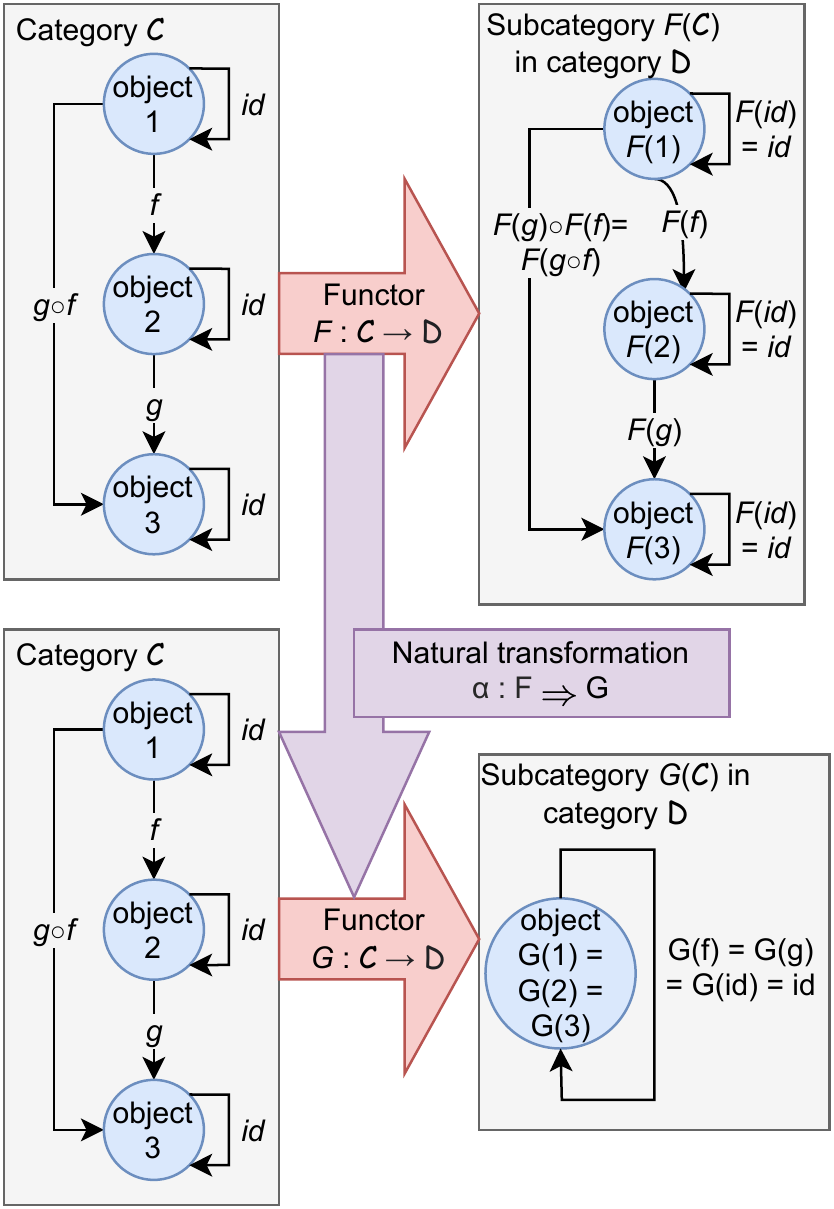}}
\caption{(a) An example of a simple functor (b) An example of a simple natural transformation $\alpha \colon F \Rightarrow G$. The component morphisms $\alpha_i \colon F(i) \to G(i)$ are defined so that they map everything to the single object in $\mathsf{D}$.}
\label{fig:category_functor_natural_transformation_examples}
\end{figure}

See \autoref{fig:category_functor_natural_transformation_examples} (b) as an example of a natural transformation.

\subsection{Kan lifts}

We discuss Kan lifts \cite{nlab:kan_lift} shortly. Kan lift is a pair consisting of a functor and a natural transformation. The problem can be expressed as a diagram
\begin{displaymath}
\begin{tikzcd}
\mathsf{A} \arrow[""{name=bar, below}]{rr}{F} \arrow[swap, dashed]{dr}{L} & & \mathsf{C} \\
& \mathsf{B} \arrow[swap]{ur}{G} \arrow[Rightarrow, to=bar, swap, "\varepsilon", shorten <=1.5ex] &
\end{tikzcd}
\end{displaymath}
where all the arrows represent functors and a natural transformation $\varepsilon \colon G \circ L \Rightarrow F$. The category theoretical problem is to find a suitable functor $L \colon \mathsf{A} \to \mathsf{B}$ and a natural transformation $\varepsilon \colon G \circ L \Rightarrow F$ which make the construction universal i.e. the natural transformation $\varepsilon$ is universal among all the suitable natural transformations which satisfy the diagram. The problem is called a \textit{lifting problem}.

\begin{definition}[Kan lift]\label{def:KanLift}
Let $F \colon \mathsf{A} \to \mathsf{C}$ and $G \colon \mathsf{B} \to \mathsf{C}$ be functors. A right Kan lift of $F$ through $G$ consists of a functor $\mathrm{Rift}_{G}F \colon \mathsf{A} \to \mathsf{B}$ and a natural transformation $\varepsilon \colon G \circ \mathrm{Rift}_{G}F \Rightarrow F$ so that they satisfy the following universal property: given any other pair of a functor and a natural transformation $(H \colon \mathsf{A} \to \mathsf{B}, \eta \colon G \circ H \Rightarrow F)$ then there exists a unique natural transformation $\gamma \colon H \Rightarrow \mathrm{Rift}_{G}F$ so that $\eta$ factors through $\varepsilon$ i.e. $\eta = \varepsilon \circ (G \circ \gamma)$. Diagrammatically if
\begin{displaymath}
\begin{tikzcd}
\mathsf{A} \arrow[""{name=bar, below}]{rr}{F} \arrow[swap, dashed]{dr}{\mathrm{Rift}_{G}F} & & \mathsf{C} \\
& \mathsf{B} \arrow[swap]{ur}{G} \arrow[Rightarrow, to=bar, swap, "\varepsilon", shorten <=1.5ex] &
\end{tikzcd}
\text{ and }
\begin{tikzcd}
\mathsf{A} \arrow[""{name=bar, below}]{rr}{F} \arrow[swap]{dr}{H} & & \mathsf{C} \\
& \mathsf{B} \arrow[swap]{ur}{G} \arrow[Rightarrow, to=bar, swap, "\eta", shorten <=1.5ex] &
\end{tikzcd}
\end{displaymath}
then
\begin{displaymath}
\begin{tikzcd}
\mathsf{A} \arrow[""{name=bar1, below}]{rr}{F} \arrow[""{name=foo, below}, bend left = 30]{dr}{H} \arrow[""{name=bar}, swap, dashed, bend right = 30]{dr}{\mathrm{Rift}_{G}F} & & \mathsf{C} \arrow[Rightarrow, from = foo, to = bar, "\gamma"] \\
& \mathsf{B} \arrow[swap]{ur}{G} \arrow[Rightarrow, to=bar1, swap, "\varepsilon", shorten <=1.5ex] &
\end{tikzcd}
\end{displaymath}
\end{definition}

The problem of finding the pair $\mathrm{Rift}_{G}F \colon \mathsf{A} \to \mathsf{B}$ and $\varepsilon \colon G \circ \mathrm{Rift}_{G}F \Rightarrow F$ is called a \textit{lifting problem}. The intuition behind Kan lifts is that we find a functor $\mathrm{Rift}_GF$ that is the best approximation which makes the triangle ''commute''. The notion of Kan lift grabs a larger collection of data transformations since we do not require that the triangle necessarily commutes in strict sense. Although the definition is abstract, we believe that is suitably flexible to describe transformations conceptually.

\subsection{Graphs}

Graphs have a three-folded role in this work. The first role of graphs is that every category is naturally a graph where objects are the vertices and morphisms are the edges. On the other hand, a graph is an abstract data model which we are formalizing in terms of category theory. Some concrete models following the graph model are property graphs and RDF graphs. The third role of graphs is that they serve as the most standard tool to model relationships in a database, for example, ER diagrams and various relational schemas are graphs. We want to emphasize that these graphs should not be confused.
\begin{definition}[Graph]\label{def:graph}
A graph $G$ is a quad $G = (V, E, src, tgt)$ where $V$ is a set of vertices, $E$ is the set of edges, $src \colon E \to V$ is the source function and $tgt \colon E \to V$ is the target function. If $e \in E$ is an edge then its source vertex is $src(e) = v$ and its target vertex is $tgt(e) = w$.
\end{definition}
When we have graphs, it is natural to talk about paths. The following notation for paths is used in \cite{functorialDataMigration}.

\begin{definition}[Path]\label{def:path}
Let $G = (V, E, src, tgt)$ be a graph. A path $p$ of length $n$ in the graph $G$ is a sequence
of connected edges in $G$. The set of all paths of length $n$ is denoted by $\mathrm{Path}_G^{(n)}$. The set of all path of $G$ is $\mathrm{Path}_G = \cup_{n \in \mathds{N}}\mathrm{Path}_G^{(n)}$.
\end{definition}

\section{Functorial instances and databases}

\subsection{Functorial representation of relational data}
We can draw a correspondence that we use categories to encode database constraints and functors to create instances. Because database instances have to follow the constraints, the structure-preserving (and thus constraint-preserving) mapping, a functor, is a natural choice to model instances and transfer constraints to them. 

David Spivak \cite{functorialDataMigration} represented a simple database definition language using categories and functors. Now we shortly recall this construction. Following his ideas, we extend relational construction to graph and hierarchical data models. When data models have their functorial representations, we can define data transformations as a solution to the lifting problem (\autoref{def:KanLift}).

\begin{definition}[Categorical path equivalence relation \cite{functorialDataMigration}]\label{def:pathEquivalence}
Let \\ $G = (V, E, src, tgt)$ be a graph. A categorical path equivalence relation, denoted by $\cong$, is an equivalence relation on the set $\mathrm{Path}_G$ of all the paths of $G$ and it has the properties listed in Definition 3.2.4 in \cite{functorialDataMigration}.
\end{definition}

We omit the full list of properties since the list is relatively long and for this work, the most important is to know that the relation $\cong$ is an equivalence relation on the set $\mathrm{Path}_G$.

\begin{definition}[Categorical schema]\label{def:categoricalSchema}
A categorical schema is $C = (G, \cong)$ where $G$ is a graph and $\cong$ is a categorical path equivalence relation on $\mathrm{Path}_G$.
\end{definition}

\begin{definition}[Schema category]\label{def:schemaCategory}
Let $C = (G, \cong)$ be a categorical schema. The schema category $\mathsf{C}$ is the category whose objects are the vertices of the graph $G$ and the morphisms are the equivalence classes of the paths of $G$ defined by $\cong$. The composition is defined as path composition with respect to the equivalence relation.
\end{definition}

The schema category consists of objects which are table descriptions, for example, similar to that we have in the ER diagram. The morphisms are induced by the foreign key constraints between the tables. Intuitively, a schema category is the category induced by the corresponding ER diagram.

\begin{definition}[Instance functor]\label{def:instanceFunctor}
Let $\mathsf{C} = (G, \cong)$ be a schema category. An instance functor $I \colon \mathsf{C} \to \mathsf{Set}$ maps the schema category to the category of sets and it satisfies the property that if $p \cong q$, then $I(p) = I(q)$.
\end{definition}

See \autoref{fig:instanceFunctorExample} (a) as an example of a relational instance functor. In \autoref{fig:instanceFunctorExample} (a) arrows are based on the constraints between the attributes. Since functional dependencies can be composed, the compositions of the dependencies are defined. A set of attributes trivially depends on itself which creates identity arrows. 

For instance, we can ask a question related to \autoref{fig:instanceFunctorExample} (a): What is the channel that the moderator with ModName \texttt{alicee} owns? The answer can be found when we compose the arrow Moderator.FollowerID $\to$ Follower.ID with the arrow Follower.OwnChannel $\to$ Channel.ID. This gives us an arrow Moderator.FollowerID $\to$ Channel.ID. The answer is the channel with id \texttt{C4} which can be read in \autoref{fig:instanceFunctorExample} (a).

The intuition behind a relational instance functor is that it sends each object $c \in \mathsf{C}$ (corresponding table description or a column in the schema) to a set $I(c) \in \mathsf{Set}$. The set $I(c)$ is the concrete instance of a table or a column. For example in \autoref{fig:instanceFunctorExample} (a), $I(\text{ChannelMods}) = \left\{(\text{C1}, \text{M1}), (\text{C2}, \text{M2}), (\text{C3}, \text{M1}), (\text{C3}, \text{M2}) \right\}$. If a morphism $f \colon c \to d \in \mathsf{C}$ corresponds a foreign key dependency between the table descriptions $c$ and $d$ in the schema, then $I(f) \colon I(c) \to I(d) \in \mathsf{Set}$ is the set valued function that sends the tuples of the table $I(c)$ to the tuples of the table $I(d)$ along the functional dependency defined by the foreign key constraint.

\subsection{Functorial representation of the graph and hierarchical data}
Bumby et al. \cite{Bumby1986} gives a category theoretical formulation for graphs. Recall that we previously defined a graph $G$ to be a quad $(V, E, src, tgt)$. Property graphs have been studied from an algebraic and category theoretical perspective already in \cite{shinavier2020algebraic}.

\begin{definition}[Graph as functor]\label{def:functorGraph}
Let $\mathsf{G}$ be the two element category which consists of the identity morphisms and two non-trivial morphisms as the diagram
\begin{displaymath}
\xymatrix{
0 \ar@/^/[r]^{\text{s}} \ar@/_/[r]_{\text{t}} & 1
}
\end{displaymath}
describes. Now a graph $G$ is a functor $G \colon \mathsf{G} \to \mathsf{Set}$ where $G(0) = E$ is the set of edges, $G(1) = V$ is the set of vertices, $G(s) \colon G(0) \to G(1) = src \colon E \to V$ is the source function and $G(t) \colon G(0) \to G(1) = tgt \colon E \to V$ is the target function. Besides, $G$ maps identity morphisms of $\mathsf{G}$ to identity functions in $\mathsf{Set}$.
\end{definition}

We do not assume that the graph would have a schema. In this sense, the construction differs from the one that we gave to the relational data. In practice, we might have a graph schema available, for example, in the cases when we are transforming relational data into graph data.

When a graph schema is available, we can encode it in the category theoretical definition. If we have a strict schema for the graph, we can generalize Spivak's approach for the relational data and assign the schema information to the objects $0$ and $1$ in \autoref{def:functorGraph}.

\begin{figure}
    \centering
     \subfigure[]{\includegraphics[width=0.49\textwidth]{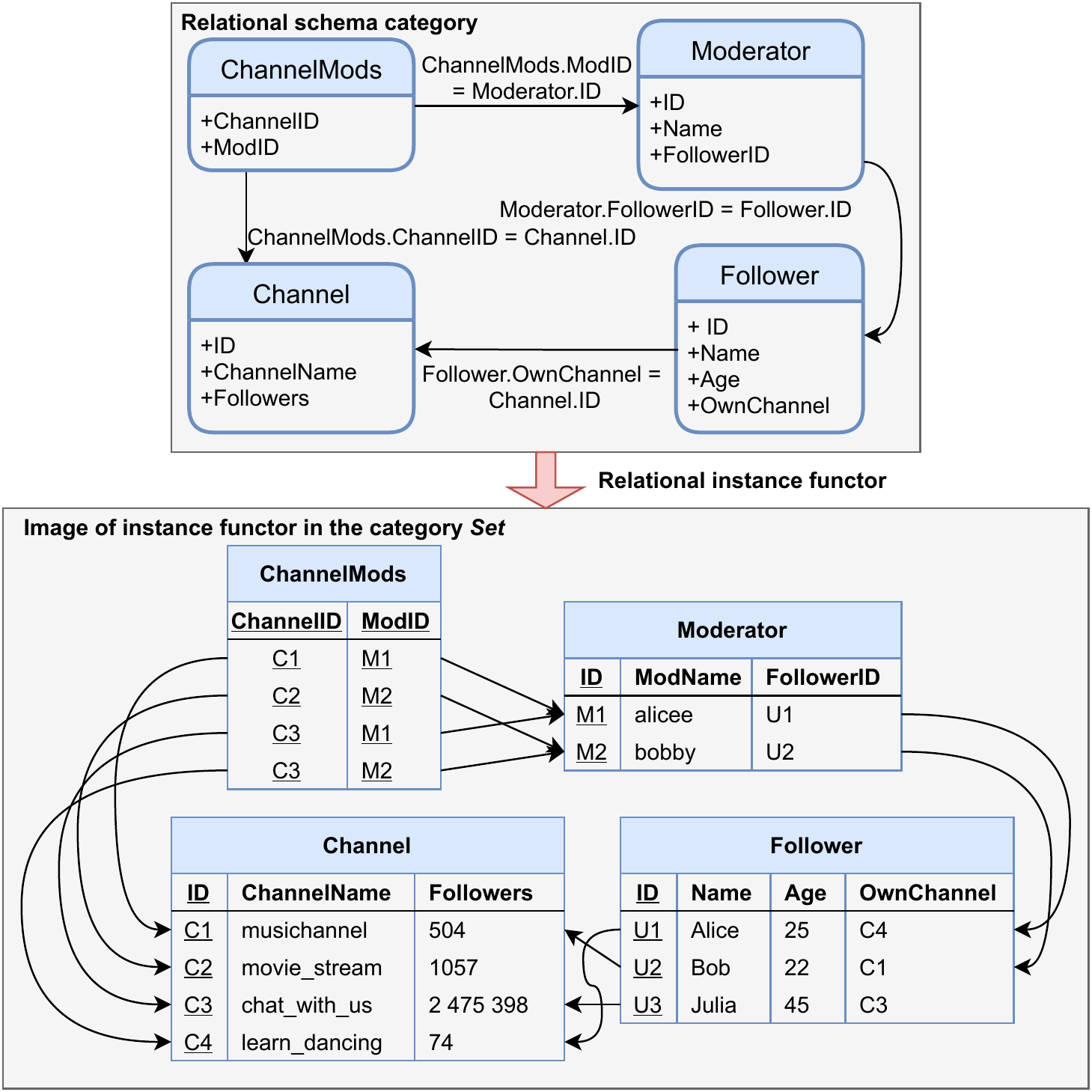}}
    \subfigure[]{\includegraphics[width=0.49\textwidth]{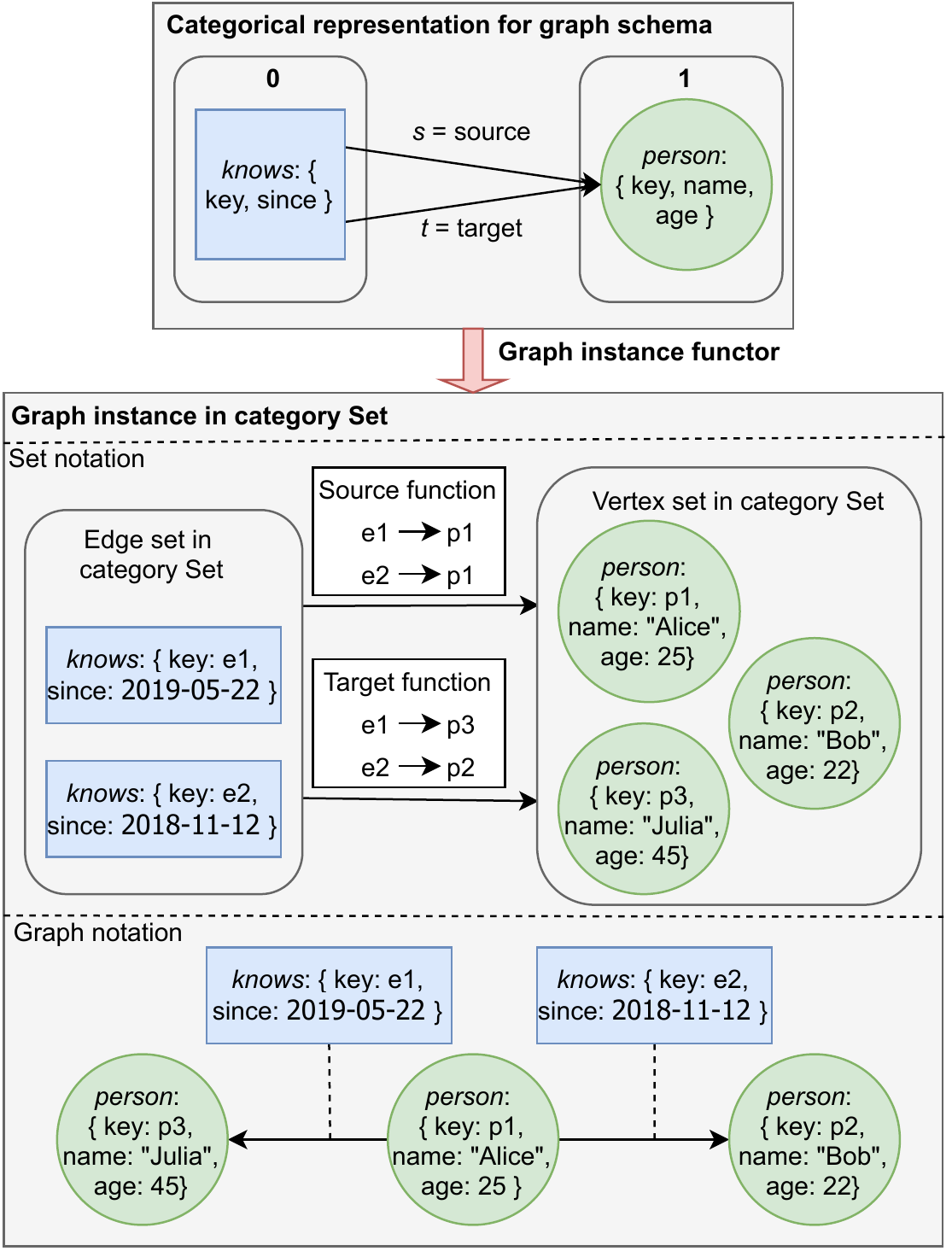}}
    \caption{(a) An example of a relational instance functor (b) The graph instance functor consists of the functor from the category that is a categorical representation for the graph schema to the category set. The graph is represented using the set notation and the conventional property graph notation.}
    \label{fig:instanceFunctorExample}
\end{figure}

The classical graph example is a social network. Let us take a property graph -oriented approach and set that the object $1$ is associated with a graph schema $(person : \left\{\text{key}, \text{name}, \text{age} \right\})$. The label \textit{person} is the label of the node and key, name and age are keys for the properties stored in nodes. For edges we can define similar structure by setting $0 = [knows : \left\{\text{key}, \text{since} \right\}]$. See \autoref{fig:instanceFunctorExample} (b) for the full construction.

As far as we know, hierarchical data, such as XML and JSON, do not have a category theoretical description that would have been studied previously. We use terms hierarchical data and tree data interchangeable. For any tree, we identify the characteristical feature that each node in the tree has exactly one parent node except the root. We can conceptually expand the tree construction so that the root is the unique node that has itself as a parent.
\begin{definition}[Tree as functor]\label{def:treeFunctor}
Let $\mathsf{T}$ be the one element category whose object is $0$ and the only non-trivial morphism is $p \colon 0 \to 0$. Diagrammatically the category is simply
\begin{displaymath}
\xymatrix{
0 \ar@(ul,ur)^{p}
}
\end{displaymath}
Now a tree is a functor $T \colon \mathsf{T} \to \mathsf{Set}$ which sends the single element $0$ to the set of nodes of the tree. The single non-trivial morphism $p \colon 0 \to 0$ is sent to the function that gives the parent node for each node in $T(0)$. If the node is the root $r$, then we define $T(p)(r) = r$. 
\end{definition}
\section{Data transformations between functorial instances}

\subsection{Intuition behind transformations represented in terms of category theory}

Before formally discussing the transformations, we show a motivating example of how the theory in the previous sections manages to unify a big part of the transformation theory.

This example is continuation to \autoref{fig:instanceFunctorExample} (b) where we had the classical social network data stored in a relational database. In our opinion, the most obvious way to store a social network is to use simple vertex- and edge-tables. The relationships are defined by foreign key constraints. The \textit{knows} table, which serves as the edge-table, has at least two foreign keys, \textit{k.personID1} and \textit{k.personID2}. These are connected to the person-table's primary key \textit{p.personID}. Diagrammatically this can be expressed as
\begin{displaymath}
\xymatrix{
\ar@/^/[r]^{\text{k.personID2 = p.personID}} \ar@/_/[r]_{\text{k.personID1 = p.personID}} \text{knows-table k} & \text{person-table p} 
}
\end{displaymath}
We note that this schema already defines a schema category (\autoref{def:schemaCategory}).

Recall the category theoretical representation for the graph in \autoref{def:functorGraph}. We can transform the relational instance into a graph in multiple ways. The first way to map the relational schema category to the graph schema category is
    \begin{displaymath}
    \text{on objects}
    \begin{cases}
    \text{p} \mapsto 1 \\
    \text{k} \mapsto 0 \\
    \end{cases}
    \text{and on morphisms}
    \begin{cases}
    (\text{p.personID  = k.personID1}) \mapsto s \\
    (\text{p.personID = k.personID2}) \mapsto t.
    \end{cases}
    \end{displaymath}
The objects $0$ and $1$ and morphisms $s$ and $t$ refer to the same objects and morphisms as in \autoref{def:functorGraph}. The second transformation is that we swap how the morphisms are mapped i.e. swap the roles of $s$ and $t$. Compared to the first transformation this inverses the direction of the edges in the resulting graph. 

Besides these two mappings, we can find two more. The third possible functor collapses the relational schema i.e. it maps everything to the object $0$ and its identity morphism:
\begin{displaymath}
\text{on objects}
\begin{cases}
\text{p} \mapsto 0 \\
\text{k} \mapsto 0 \\
\end{cases}
\text{and on morphisms}
\begin{cases}
(\text{p.personID  = k.personID1}) \mapsto \text{id}_0\\
(\text{p.personID = k.personID2}) \mapsto \text{id}_0.
\end{cases}
\end{displaymath}
The fourth possible functor is similar to the previous functor but maps everything to the object $1$.
The benefit of the category theoretical formulation for transformations is that we can mathematically characterize, that the transformation which sends the knows-table to vertices and person-table to edges, is not valid because such transformation is not a functor.
   
The transformations $3.$ and $4.$ have problems although they are well-defined functors. Thus functoriality is not a sufficient condition to characterize meaningful transformations. It does not make sense to map everything to edges (the result of the transformation $3.$) because a valid edge needs to have a source and a target vertex. Also, a graph that contains only vertices without edges (the result of the transformation $4.$) is not meaningful because edges are necessary for the most important graph operations. Thus we require that the functor should be \textit{full} (\autoref{def:functor}) to be relevant in practice. As we see, the transformations $3.$ and $4.$ as functors are not full but transformations $1.$ and $2.$ are.

\subsection{Data transformation as lifting problem}

Data and schema transformations are usually modeled as mappings from a source database to a target database. We base our data and schema transformation on Kan lifts \cite{nlab:kan_lift}. Lifting problems have been considered in database theory also previously in \cite{Spivak2013DatabaseQA}. As \autoref{def:KanLift} shows, the lift consists of two components: a functor and a natural transformation. Informally, the functor part is a schema mapping which describes a set of rules which define how the data items are mapped at a schema level. The functor is required to be \textit{full} (\autoref{def:functor}) because functors which are not full are not practically meaningful as the discussion in the previous section shows. Along the functor, we have a natural transformation which is data mapping. The pair satisfies the universal property which creates certain classification for transformations. The nature of this classification is still an open question.

The category theoretical approach to data and schema transformations reveals a crucial problem in transformation research. The problem is the separation of data and schema. In a world where relational databases are still the dominant databases, the division of data and schema is obvious. But the problem is apparent with the schemaless or schema-free models such as graphs and documents. If graph and document data transformations are approached from the relational perspective, we are likely to face problems. With category theory, we can model as much structure as the data has. Modeling transformations as pairs of mappings describes transformations more rigorously than a single total function between data sets.

Let $I_1 \colon \mathsf{C_1} \to \mathsf{Set}$ and $I_2 \colon \mathsf{C_2} \to \mathsf{Set}$ be two data instances as functors where the functors can represent either relational, graph or hierarchical instance functors as described in Definitions \ref{def:instanceFunctor}, \ref{def:functorGraph}, and \ref{def:treeFunctor}. The question is how do we generally find a transformation between the data instances $I_1$ and $I_2$. The problem can be expressed as a diagram
\begin{displaymath}
\begin{tikzcd}
\mathsf{C_1} \arrow{rr}{I_1} \arrow[swap, dashed]{dr}{F} & & \mathsf{Set} \\
& \mathsf{C_2} \arrow[swap]{ur}{I_2} &
\end{tikzcd}
\end{displaymath}
where the functor $F \colon \mathsf{C_1} \to \mathsf{C_2}$ is the schema transformation mapping between the categorical representations of the schema categories $\mathsf{C_1}$ and $\mathsf{C_2}$. The second part of the transformation consists of a natural transformation $\varepsilon \colon I_2 \circ F \Rightarrow I_1$ which obeys certain laws. If we assume that we have the two diagrams
\begin{displaymath}
\begin{tikzcd}
\mathsf{C_1} \arrow[""{name=bar, below}]{rr}{I_1} \arrow[swap, dashed]{dr}{F} & & \mathsf{Set} \\
& \mathsf{C_2} \arrow[swap]{ur}{I_2} \arrow[Rightarrow, to=bar, swap, "\varepsilon", shorten <=1.5ex] &
\end{tikzcd}
\text{ and }
\begin{tikzcd}
\mathsf{C_1} \arrow[""{name=bar, below}]{rr}{I_1} \arrow[swap, dashed]{dr}{H} & & \mathsf{Set} \\
& \mathsf{C_2} \arrow[swap]{ur}{I_2} \arrow[Rightarrow, to=bar, swap, "\eta", shorten <=1.5ex] &
\end{tikzcd}
\end{displaymath}
where the second diagram has a functor $H \colon \mathsf{C_1} \to \mathsf{C_2}$ and $\eta \colon I_2 \circ H \Rightarrow I_1$ a natural transformation. We then require that there exists a \textit{unique} natural transformation $\gamma \colon H \Rightarrow F$ such that  $\eta = \varepsilon \circ (I_2 \circ \gamma)$.

\begin{definition}[Data and schema transformation]\label{def:dataSchemaTransformation}
Let $I_1 \colon \mathsf{C}_1 \to \mathsf{Set}$ and $I_2 \colon \mathsf{C}_2 \to \mathsf{Set}$ be two data instances. A transformation from $I_1$ to $I_2$ is a Kan lift $(\mathrm{Rift}_{I_2}I_1 \colon \mathsf{C_1} \to \mathsf{C_2}, \ \varepsilon \colon I_2 \circ \mathrm{Rift}_{I_2}I_1 \Rightarrow I_1)$ so that the functor $\mathrm{Rift}_{I_2}I_1$ is a full functor.
\end{definition}

\begin{figure}
    \centering
    \includegraphics[scale=0.5]{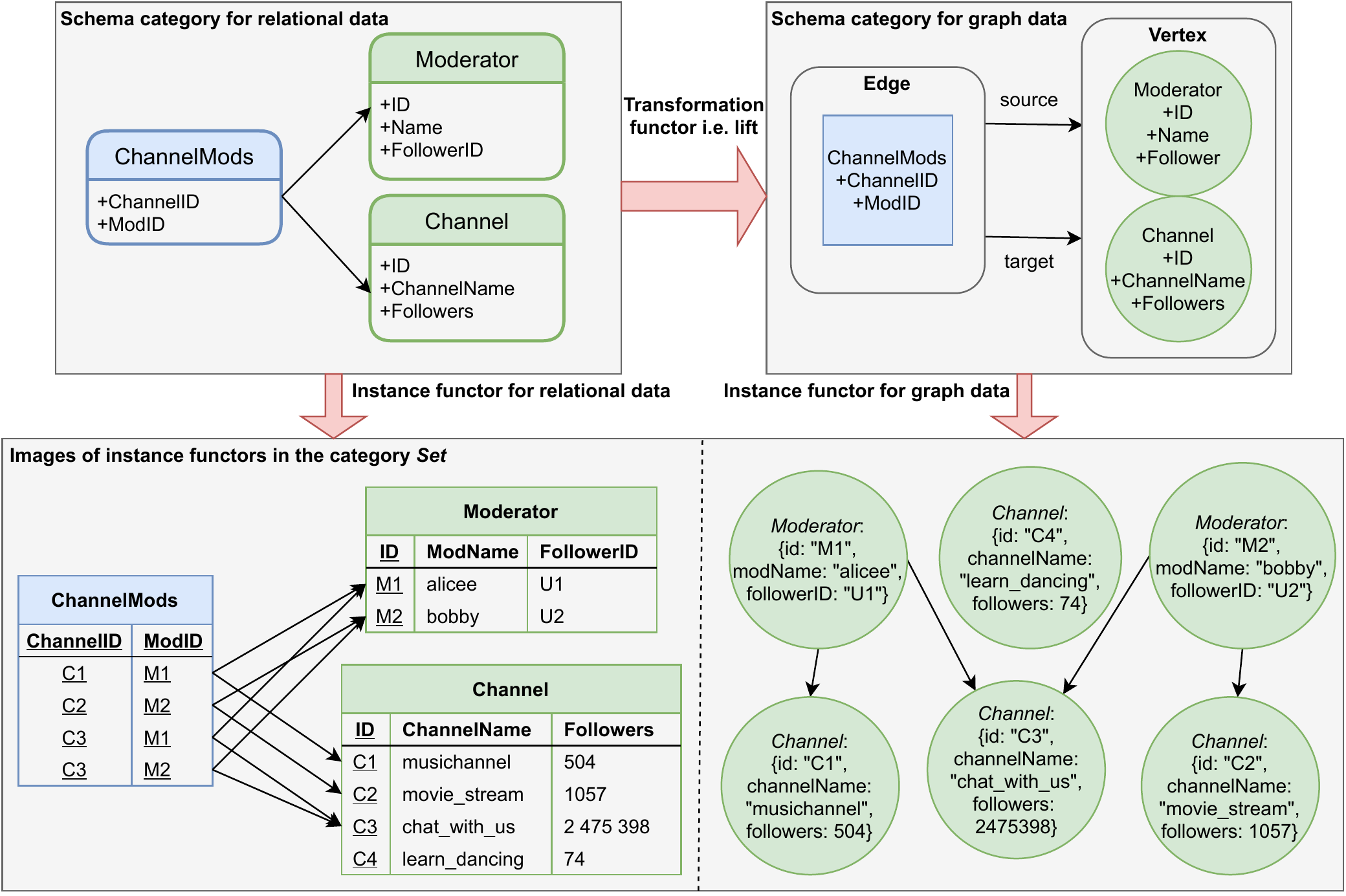}
    \caption{Example transformation from relational to property graph.}
    \label{fig:relational2propertygraph}
\end{figure}

We recall our example relational database instance in \autoref{fig:instanceFunctorExample} (a). In order to transform the relational instance to a property graph we need to construct a functor from the relational schema category to the graph schema category and define the natural transformation. \autoref{fig:relational2propertygraph} describes the full transformation and the coloring codes the corresponding elements in each category. Informally, the natural transformation in the example could be understood so that for each object in the relational schema category, we have a mapping that tells how the corresponding relational data object in the category Set is mapped to the graph data object in the category Set. 

\section{Conclusions and future work}
When the variety and amount of data grows, the need for polystores and multi-model databases is urgent. The efficient utilization of the systems requires a precise theory of how the systems operate and how they are modeled. So far, there has been extensive research on practical and implementational aspects. Without a proper theoretical framework, the field is left scattered. We are answering this challenge by formalizing the three most common data models and the data and schema transformations between them. We continued previous research and contributed by formalizing graph and hierarchical models functorially. We then focused on data and schema transformations between the functorial instances. Kan lifts require more studying as a basis for transformations but it seems a promising direction. 

Query transformations form another half of the transformation systems. A query can be transformed correctly if the data is transformed correctly. This ties both transformations together which makes the modeling challenge still harder. Future work would include formalizing and unifying query transformations. In the case of SQL, the topic has already been studied in \cite{Spivak2013DatabaseQA}.

We identify that there is a need to model temporal data better. The problem of temporality is rarely addressed in polystore, multi-model database, and transformation research. Usually, the implicit assumption, especially in transformation frameworks, is that the systems are dealing with static data. Of course, that is hardly ever true and data changes and expands constantly. We believe that with category theory we can naturally include a time component to data. 

\section*{Acknowledgement}
This paper is partially supported by Finnish Academy Project 310321 and Oracle ERO gift funding.

\bibliographystyle{splncs04}
\bibliography{ref}

\end{document}